\documentstyle[aps,prb,preprint,graphicx,ifthen]{revtex}

\newcommand{\material}{Sr$_2$CuO$_2$Cl$_2$}
\newcommand{\lacp}{La$_2$CuO$_4$}
\newcommand{\co}{CuO$_2$}
\newcommand{\op}[1]{{#1}}
\newcommand{\Op}[1]{\bar{#1}}
\renewcommand{\vec}[1]{{\bf #1}}
\newcommand{\vk}{\vec{k}}
\newcommand{\vq}{\vec{q}}
\newcommand{\xc}[3]{\op{a}^{#1\dagger}_{#2,#3}\,}
\newcommand{\xa}[3]{\op{a}^{#1}_{#2,#3}\,}
\newcommand{\Xc}[4]{\Op{a}^{#1\dagger}_{#2,#3}(#4)\,}
\newcommand{\Xa}[4]{\Op{a}^{#1}_{#2,#3}(#4)\,}
\newcommand{\E}[3]{E^{#1}_{#2,#3}\,}
\newcommand{\omlo}{\omega_{0}}
\newcommand{\bq}{\op{b}_{\vq}\,}
\newcommand{\bcq}{\op{b}^\dagger_{\vq}\,}
\newcommand{\bcmq}{\op{b}^\dagger_{-\vq}\,}
\newcommand{\coup}[3]{\ifthenelse{\equal{#3}{}}%
  {M^{#1}_{#2}\,}%
  {M^{#1}_{#2,#3}\,}}
\newcommand{\om}{\omega}
\newcommand{\G}[4]{G^{#1}(#2#3;#4)\,}
\newcommand{\GO}[4]{G^{#1}_{0}(#2#3;#4)\,}
\newcommand{\W}[5]{\ifthenelse{\equal{#2}{}}%
   {W_{#1}\,}%
   {W^{#2}_{#1}(#3#4;#5)\,}}
\newcommand{\F}[5]{\ifthenelse{\equal{#3}{}}%
   {F_{#1}\,}%
   {F^{#2}_{#1}(#3#4;#5)\,}}
\newcommand{\step}[1]{\theta(#1)\,}
\newcommand{\Int}[3]{\int\limits_{#1}^{#2} \!\!d#3\,}
\newcommand{\expec}[1]{\left< #1 \right>}
\newcommand{\abs}[1]{\left|#1\right|}

\newcommand{\fbar}[2]{\bar{f}(#1, #2)\,}
\newcommand{\bose}{N_0\,}
\newcommand{\eps}{\varepsilon}
\newcommand{\mxx}{M_{xx}}
\newcommand{\mxc}{M_{xc}}
\newcommand{\mxd}{M_{xd}}
\newcommand{\mcc}{M_{cc}}
\newcommand{\mcd}{M_{cd}}
\newcommand{\dos}[2]{z_{#1}(#2)\,}

\def\dx2y2{$d_{x^2 - y^2}$}

\begin{document}
\draft
\title{Evidence of phonon mediated coupling between charge transfer and ligand
field excitons in \material}

\author{R.~L\"ovenich, A.~B.~Schumacher\cite{schuh_add}, J.~S.~Dodge\cite{dodge_add}, and D.~S. Chemla}
\address{Department of Physics, University of California at
   Berkeley, Berkeley, California 94720}
\address{Materials Sciences Division, E.~O.~Lawrence Berkeley
   National Laboratory, Berkeley, California 94720}

\author{L.~L.~Miller}
\address{Ames Laboratory and Department of Physics, Iowa State
   University, Ames, IA 50011}

\date{\today}

\maketitle

\begin{abstract}
We present a comparative experimental and theoretical
investigation of the two--dimensional charge-transfer gap in the
strongly correlated material \material. We observe an Urbach
behaviour in the absorption profile over a surprisingly wide range
of energies and temperatures. We present a model that accounts for
phonon scattering to infinite order and which allows us to explain
the data accurately by assuming coupling of the charge transfer
gap exciton to lower energy electronic excitations.
\end{abstract}

\newpage

\section{Introduction}

The excitation spectrum of transition metal oxides is
characterized by strong coupling of spin, charge, orbital and
lattice degrees of freedom, and the complexity of these
interactions has severely hindered progress in understanding the
phenomenon of high--temperature superconductivity in the layered
copper oxides.  Even the basic energy level spectrum of the
undoped, insulating cuprates continues to be controversial.  The
dominant optical absorption feature in these materials is a broad
excitonic peak at the charge-transfer (CT) gap, about half an
electron volt wide, which exhibits strong dependence on
temperature and doping~\cite{uchida91,falck92}. Despite the strong
interest in insulating cuprates and the fundamental physical
significance of the CT gap exciton, few attempts have been made to
explain its structure and coupling to
phonons\cite{falck92,Zhang98}.

In this paper, we present an experimental investigation of the
linear absorption of \material\ and a new theory for the CT
excitation in the CuO-plane. The theory includes the coupling of
the exciton to phonons and to an additional, low energy electronic
continuum of states. It explains the linear absorption of
\material, and in particular the Urbach behaviour over a wide
range of temperatures and energies.

\section{Sample structure}

\material~is a two-dimensional, spin-$1/2$ Heisenberg
antiferromagnet with a N\'{e}el temperature T$_{N}\approx$ 250~K
and an exchange interaction energy J=125~meV~\cite{tokura90}. In
the ground state a single hole per \co\ unit cell is located on
the Cu site and exhibits a 3\dx2y2\  symmetry.

The lowest energy optically allowed electronic excitation in this
material corresponds to the Cu~3\dx2y2\ to O~2$p\sigma$ transfer
of a hole in the \co~layer. This transition classifies the
material as a charge--transfer insulator \cite{zaanen85}. The
optical absorption spectrum consists mainly of a rather sharp peak
near the band edge at about 2~eV and another wider peak at about
2.5~eV \cite{choi99}. A detailed analysis of the energy region
below the band edge reveals a weak absorption step at about
1.4~eV, two orders of magnitude smaller than the main CT
structure. It is attributed to a Cu ligand field transition
\cite{perkins93}, which should be optically forbidden, but
acquires a small oscillator strength owing to a breaking of the
crystal symmetry. This transition has also been observed in
resonant X-ray Raman spectroscopy\cite{kuiper98} and large--shift
Raman scattering\cite{salamon95}.

The dominant peak at the band edge is attributed to a bound
excitonic state. A model by Zhang and Ng \cite{Zhang98} predicts
an exciton formed by an electron at a Cu site and a delocalized
hole at the four neighbouring O sites, with a stronger weight at
the O site from which the electron came. The model gives a large
bandwidth for the center of mass motion of this exciton of about
1.5~eV\cite{wang96}. This means that the exciton has an effective
mass smaller than the electron or hole alone, which can be
understood by the fact that the transfer of the exciton from one
site to another does not disturb the antiferromagnetic order of
the background.

\section{Experiment}

\begin{figure}
     \begin{center}
       \includegraphics[width=0.85\columnwidth]{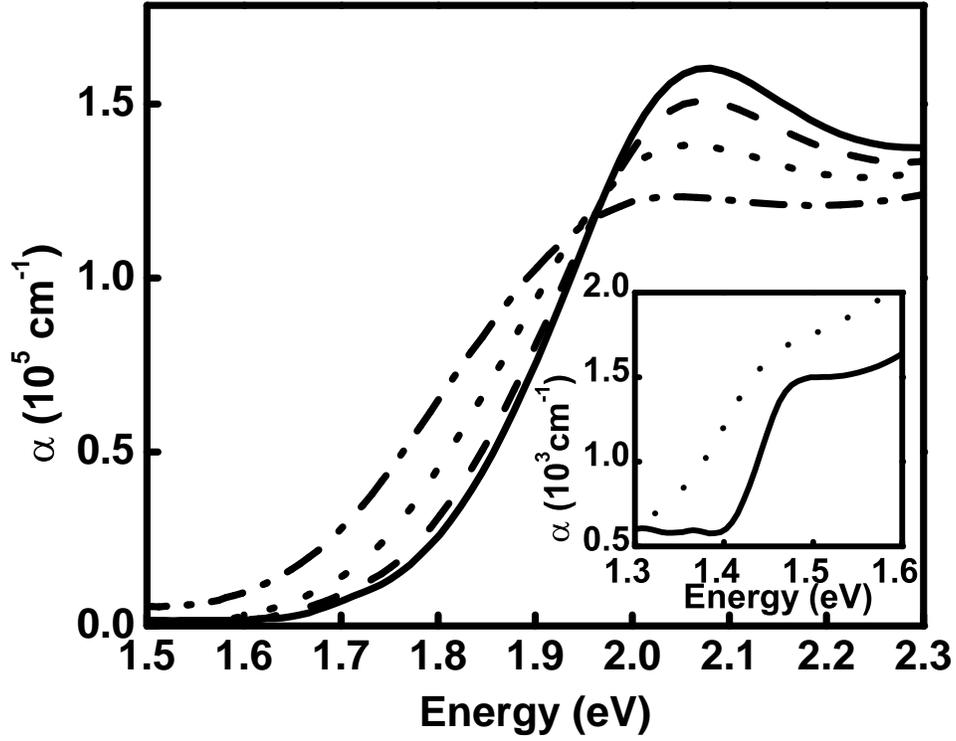}
         \caption{Absorption coefficient of \material~at different
           temperature, measured on a l=950~\AA~thick sample. Inset
         (l=300$\mu$m): the weakly allowed d-d transition. Solid lines
         T=15~K, dashed line T=150~K, dotted lines T=250~K,
         dashed-dotted line T=350~K}
         \label{linear absorption}
     \end{center}
\end{figure}

We used high quality single crystals of \material~which were grown
by the technique described in Ref.~\onlinecite{Miller90}. This
method yields platelets with a fairly large (001) surface. By
tape-cleaving the crystals, we reduced the sample thickness, $d$,
to below 100~nm. These thin samples enabled us to measure the
absorption coefficient, $\alpha(\omega)$, across the charge
transfer gap directly, by comparing as a function the frequency
$\omega$ the intensity transmitted through the sample on the
substrate, $I_{Tot}$, with the intensity transmitted through the
substrate only, $I_{Sub}$: $\alpha(\omega) =
-\frac{1}{d}\,\ln(\frac{I_{Tot}}{I_{Sub}})$. This procedure is
more precise and direct than the previous reflectivity
measurements, which relied on Kramers--Kronig transformation to
extract information about the absorption
coefficient\cite{zibold96}. Figure~\ref{linear absorption} shows
the linear absorption coefficient for selected temperatures. The
crystal used for this measurement was $d=95\pm$5~nm~thick and held
inside a cold finger cryostat. The inset details the weak onset of
the absorption at 1.4~eV and was measured using a 300~$\mu$m~thick
sample.

\begin{figure}
     \begin{center}
        \includegraphics[width=0.85\columnwidth]{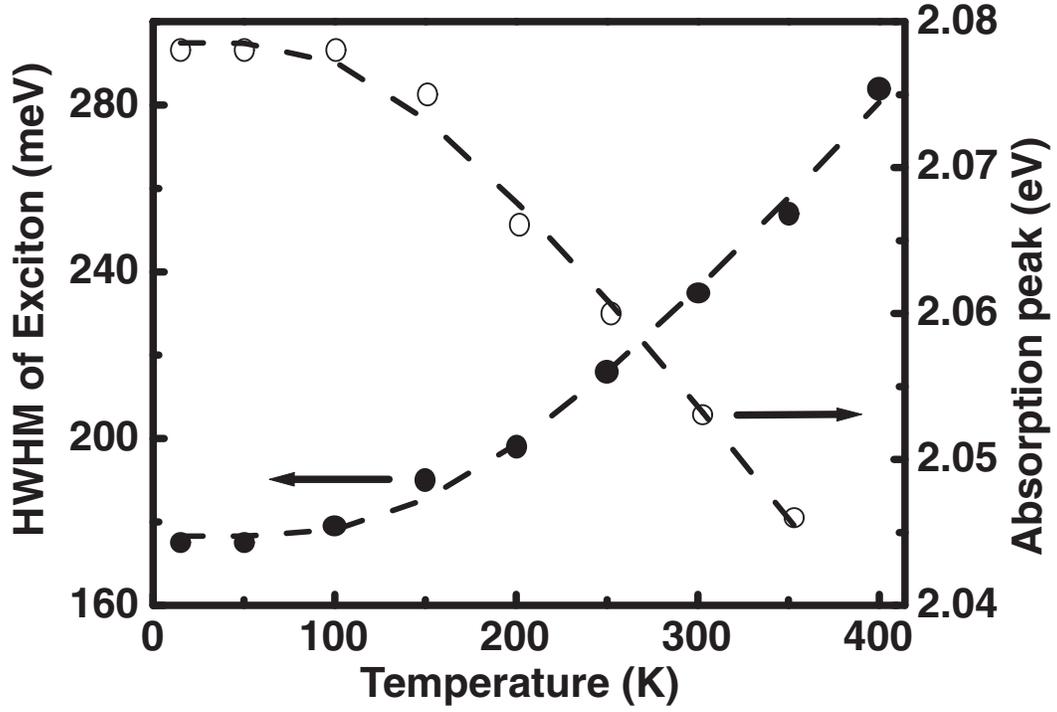}
         \caption{Half width (solid circles, left axis) and energy
           (open circles, right axis) of the absorption peak at
           different temperatures. The dotted shows the best fit to a
           single oscillator Bose--Einstein occupation function with
           E$_{boson}$=45~meV}
         \label{Shift}
     \end{center}
\end{figure}

\begin{figure}
     \begin{center}
        \includegraphics[width=0.85\columnwidth]{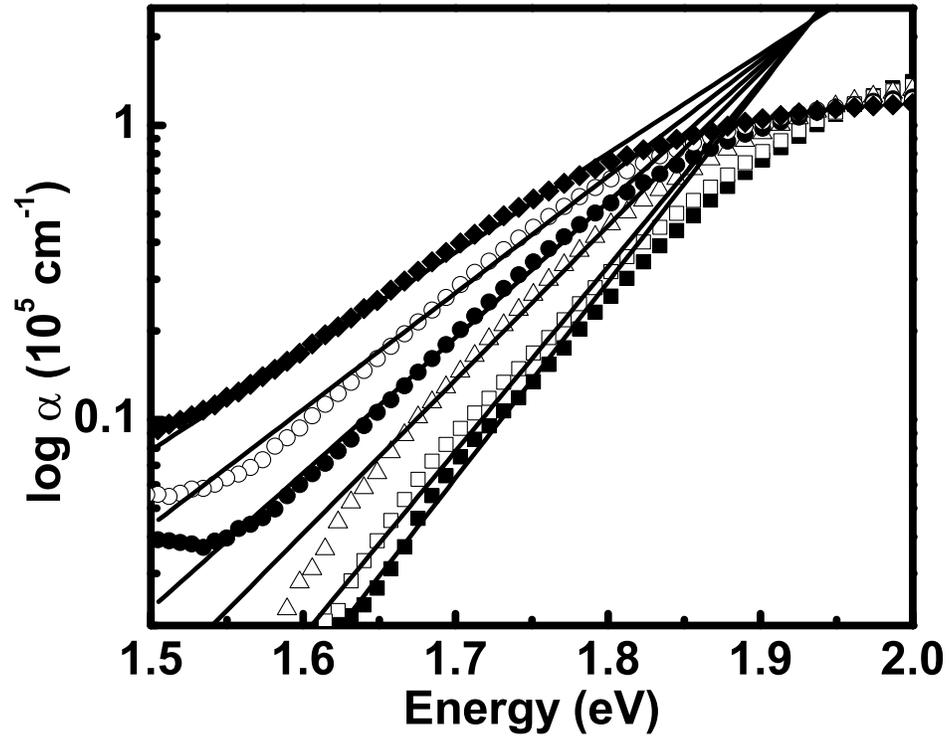}
         \caption{Logarithmic plot of the absorption coefficient of
           \material~at selected temperatures, displaying the
         Urbach--tail. Solid squares T=15~K, open squares T=~150~K,
         open triangles T=250~K, solid circles T=300~K, open circles
         T=350~K, diamonds T=400~K. The solid lines are fits to the
         Urbach formula.}
         \label{Urbach Tail}
     \end{center}
\end{figure}

The excitonic peak shows a strong temperature dependence. With
increasing temperature the width --- measured as the half width at
half maximum --- increases from 180~meV at 15~K to 270~meV at
350~K. The maximum exhibits a redshift of about 80~meV in this
temperature range. The temperature dependence of the excitonic
peak is shown in figure~\ref{Shift}. Both the shift towards lower
energies and the broadening can be fit by a Bose--Einstein
occupation function with a single oscillator\cite{falck92}. From
this fit we find the energy of this oscillator,
$\hbar\omlo\approx45$~meV, to agree with that of a previously
identified bond--bending longitudinal optical (LO)
phonon\cite{Tajima91} and well below the characteristic magnetic
energy of J=125~meV in \material. Experimental evidence therefore
suggests coupling of the CT-exciton to phonons rather than to
magnetic excitations as the origin of its red shift and broadening
\cite{falck92}. This is further supported by a careful analysis of
the line shape of the low energy side of the CT excitation. As
shown in figure~\ref{Urbach Tail}, $\alpha(\omega)$ exhibits an
exponential behaviour over a wide range of temperatures and
energies. We find the experimental data can be fit very well to
the Urbach formula
\begin{eqnarray}
   \alpha(\omega,T) & = &
   \alpha_{0}\exp{\frac{\sigma(T)(E-E_{0})}{k_{B}T}};\\
   \sigma(T) & = &
   \sigma_{0} \,\frac{2k_{B}T}{\hbar\omlo} \,
      \tanh{\frac{\hbar\omlo}{2k_{B}T}}.
\end{eqnarray}
where E$_{0}$ and $\alpha_{0}$ are temperature independent
parameters, and $\hbar\omlo$ is the energy of the phonon to which
the exciton couples. The parameter $\sigma_{0}$ yields information
on the strength of this coupling and E$_{0}$ is related to the
strength of the exciton binding. From a global best fit to the
data plotted in figure~\ref{Urbach Tail} data we obtain
$\hbar\omlo$=45~meV, E$_{0}$=1.95~eV and $\sigma_{0}=0.35$. The
latter indicates a very strong coupling between excitons and
phonons. Furthermore, the fact that E$_{0}$ is smaller than the
energy E($\alpha$=max) where the absorption is maximal, suggests a
tightly bound exciton~\cite{kurik71}.

It is worth noting that it is very surprising to observe an Urbach
behavior at temperature as low as 15~K where the occupation
probability of LO--phonons with $\hbar\omlo$=45~meV is
$N_0\approx$2$\times$10$^{-16}$ and absorption of thermal phonons
is negligible.

\section{Theory}

To the best of our knowledge, the model of Falck et
al.\cite{falck92} is the only one in the literature describing the
temperature dependence of the CT excitations. Falck et~al.
\cite{falck92} investigated the absorption line shape of a related
compound, the charge--transfer insulator \lacp. The excitonic peak
in this material also shows a temperature shift and broadening.
Approximating the Coulomb interaction with an attractive contact
interaction between the electron and hole, they derived a simple
expression for the absorption coefficient. While without
interactions the interband contribution to the absorption is just
the density of states, the Coulomb interaction leads to an
increase of the absorption near the band edge. This describes the
basic shape of the spectrum. Falck et~al.\  account for the shift
of the band edge and the finite lifetime of the exciton by
inserting the real and imaginary part of the one--phonon self
energy. With four parameters they are able to fit well the linear
absorption in \lacp.

Our experimental results show the exponential behavior of an
Urbach tail down to very low temperatures. It is in striking
contradiction with the prediction of Ref.~\onlinecite{falck92},
which by construction always produces a Lorentzian tail. The
coupling constant used in Ref.~\onlinecite{falck92} was $\alpha =
10.8$\cite{coupling}. This is an extremely high value compared to
normal semiconductors (where it is well below 1.0) and even other
materials with a strong electron--phonon coupling. For such a
high coupling constant the one--phonon approximation is no longer
valid.

Since a strong coupling to LO--phonons seems to be the reason for
the strong temperature dependence of the spectra, we propose a
more advanced model, taking into account many--phonon processes.
The approach was inspired by results obtained for conventional
semiconductors \cite{liebler85}, where Urbach tails are found for
high temperatures.

The theoretical analysis is organized as follows: First we
introduce the Hamiltonian in an exciton basis. The Green's
function of that Hamiltonian will be calculated by means of the
cumulant expansion. We will then discuss the dependence of the
results on model parameters for different cases of physical
interest. We will show that it is necessary to include ligand
field (LF) excitations, which were not yet known at the time of
Ref.~\cite{falck92}. These are the states into which the CT
exciton can scatter even at low temperatures by phonon emission,
thus leading to the Urbach tail.

The model Hamiltonian:
\begin{eqnarray}
   \label{hamiltonian}
   H & = & H_{0} + V
   \\
   \label{h0-def}
   H_{0} & = &\sum_{jn\vk} \E{j}{n}{\vk} \xc{j}{n}{\vk} \xa{j}{n}{\vk}
     + \hbar \omlo \sum_{\vq} \bcq \bq
   \\
   V & = &\sum_{jj'}\sum_{nm\vq} \coup{jj'}{nm}{\vq}
       \xc{j}{n}{\vk+\vq} \xa{j'}{m}{\vk}
       \Big[ \bq + \bcmq \Big]
   \label{v-def}
\end{eqnarray}
is written in an exciton basis, where $\xc{j}{n}{\vk}$ is the
creation operator for an exciton of type $j$, internal quantum
number $n$ and center of mass momentum $\vk$. In the case of the
$j$=CT exciton we account for a bound state ($n=0$), and for the
unbound exciton electron--hole continuum ($n>0$). The pair
$(n,\vk)$ can then be transformed to the pair of electron and hole
momentum, which is more suited to describe the free electron--hole
pair. In the case of the ligand field excitation ($j$=LF) we
consider continuum states only. The operators $\bq$ and $\bcq$
respectively describe the annihilation and creation of an
dispersionless LO--phonon of frequency $\omlo$.

The imaginary part of the dielectric function and hence the
spectral features of the absorption coefficient are given by the
imaginary part of the retarded Green's function
\begin{equation}
   \label{absorption-greens}
   \varepsilon_{2}(\om) \propto \mbox{Im}
     \bigg\{ \sum_{j=\text{CT},n\vk} \G{j}{n}{\vk=0}{\om} \bigg\}
   \; ,
\end{equation}
where the summation includes only the optically active CT-exciton.

Since the excitation starts from a ground state
characterized by empty hole and electron states, the retarded
Green's function equals the time--ordered one, for which an
expansion in the electron--phonon coupling exists:
\begin{eqnarray}
   \label{g-expansion}
   \G{j}{n}{\vk}{t} & = & -\frac{i}{\hbar} \step{t}
     \sum_{\nu=0}^{\infty} \W{\nu}{j}{n}{\vk}{t}
   \\
   \label{w-def}
   \W{\nu}{j}{n}{\vk}{t} & = &
   \frac{(-i)^{2\nu}}{\hbar^{2\nu}(2\nu)!}
   \Int{0}{t}{t_1} \dots \Int{0}{t}{t_{2\nu}}
   \expec{T \Xa{j}{n}{k}{t} \Op{V}(t_1) \dots \Op{V}(t_{2\nu})
     \Xc{j}{n}{k}{0} }
   \;,
\end{eqnarray}
with the time-ordering operator $T$. The unperturbed Green's
function is given by $\GO{j}{n}{\vk}{t} = -i/\hbar \, \step{t}
\exp(-i\E{j}{n}{\vk} t/\hbar)$. The first order term in
(\ref{g-expansion}) would give rise to one--phonon processes.
However, it is possible to account for multi-phonon processes by
resumming the expansion Eq.~(\ref{g-expansion}) in form of an
exponential, leading to the cumulant expansion or linked--cluster
theory. We write
\begin{eqnarray}
   \label{g-resummation}
   \G{j}{n}{\vk}{t} = \GO{j}{n}{\vk}{t}
   \exp\bigg( \sum_{\nu=1}^{\infty} \F{\nu}{j}{n}{\vk}{t} \bigg)
   \;,
\end{eqnarray}
where $\F{\nu}{}{}{}{}$ denotes a contribution which contains the
$2\nu$-th power of the coupling matrix--element. Comparing the
different powers of the expansions (\ref{g-expansion}) and
(\ref{g-resummation}) the following relations hold:
\begin{eqnarray}
   \label{F1-def}
   \F{1}{j}{n}{\vk}{t} & = &
     e^{i\E{j}{n}{\vk} t/\hbar}\, \W{1}{j}{n}{\vk}{t}
   \\
   \label{F2-def}
   \F{2}{j}{n}{\vk}{t} & = &
     e^{i\E{j}{n}{\vk} t/\hbar} \W{2}{j}{n}{\vk}{t}
     - \frac{1}{2!}\, \F{1}{j\,2}{n}{\vk}{t}
   \;.
\end{eqnarray}
The advantage of this re-summation is that $F_{1}$ already
includes independent phonon scattering to infinite order. Thus as
long as they are independent from each other, the first term in
the cumulant expansion $F_{1}$ accounts for all these
contributions in contrast to the expansion (\ref{g-expansion}).
Similarly $F_{2}$ describes all electron--phonon interactions
involving simultaneously two phonons, etc. The cumulant expansion
(\ref{g-resummation}) is expected to converge much faster than
(\ref{g-expansion}) and to be appropriate for low and intermediate
coupling constants. Up to now we restricted  ourself to the lowest
order only, which means that we treat the fields induced by the
lattice distortions classically.

The time--integrations in (\ref{w-def}) can be performed
analytically, yielding:
\begin{eqnarray}
   \F{1}{j}{n}{\vk=0}{t} & = &
   \sum_{h'm\vq} \abs{\coup{jj'}{nm}{\vq}}^2 \Big\{
   (\bose + 1) \fbar{\E{j'}{m}{\vq} - \E{j}{n}{0} + \hbar \omlo}{t}
   \nonumber \\
   & &
   + \bose     \fbar{\E{j'}{m}{\vq} - \E{j}{n}{0} - \hbar \omlo}{t}
   \Big\}
   \;.
   \label{F1-summation}
\end{eqnarray}
The time dependence is described by the function
\begin{equation}
   \label{fbar-def}
   \fbar{\eps}{t} = \frac{1}{\eps^2}\,
   \big( 1 - e^{-i\eps t/\hbar} - \frac{i\eps t}{\hbar} \big)
   \;.
\end{equation}
This function is regular for all values of $\eps$ and has its
maximum value at $\eps=0$ for finite times, which in
Eq.~(\ref{F1-summation}) corresponds to $\E{j'}{m}{\vq} =
\E{j}{n}{0} \pm \hbar \omlo$, i.e.\ the absorption and emission of
LO-phonons in the $(j', m \vq) \leftrightarrow (j, n )$
transition. At low temperatures only the emission term
proportional $(\bose+1) \rightarrow 1$ contributes, which peaks
out one phonon energy below the CT exciton.

In order to keep our model simple, we neglect the $\vq$ dependence
of the coupling matrix elements $\coup{jj'}{nm}{\vq}$ as well as
the dependence on the relative momentum if $m>0$ or $n>0$, i.e.,
for the electron--hole continuum. This means we have to consider
the following five scattering processes:
\begin{itemize}
\item[$\mxx$] Scattering from the CT-exciton into its center
   of mass continuum;
\item[$\mxc$] Scattering from the CT-exciton into its own
   electron--hole continuum;
\item[$\mcc$] Scattering within the CT electron--hole continuum;
\item[$\mxd$] Scattering from the CT-exciton into the LF exciton band;
\item[$\mcd$] Scattering from the CT electron--hole continuum into
   the LF exciton band;
\end{itemize}

Since the matrix--elements do not depend on momentum variables any
more, the summations over momentum variables in (\ref{F1-summation})
can be transformed into energy integrals with the density of states
into which the CT-exciton is scattered:
\begin{eqnarray}
   \label{F1-energy}
   \F{1}{}{n=\{x,c\},}{0}{t} & = &
   \Int{}{}{\eps}
   \sum_{m=\{x,c,d\}} \coup{}{nm}{} \dos{m}{\eps}
   \\
   & \times &
   \Big\{
     (\bose + 1) \fbar{\eps - \E{}{n}{0} + \hbar \omlo}{t}
     + \bose \fbar{\eps - \E{}{n}{0} - \hbar \omlo}{t}
   \Big\}
   \;.
   \nonumber
\end{eqnarray}
The densities of states are defined as:
\begin{eqnarray}
   \label{zx-def}
   \dos{x}{\eps} & = &
     \frac{1}{N} \sum_{\vq} \delta(\eps - \E{}{x}{\vq})
   \hspace{1em} \mbox{and}
   \\
   \label{zc-def}
   \dos{c}{\eps} & = &
      \frac{1}{N^2} \sum_{\vq\vk} \delta(\eps - \E{}{\vk}{\vq})
   \;.
\end{eqnarray}

We model the energy dispersion of the exciton by a two dimensional
tight binding dispersion $\E{}{x}{\vq} = B/4 \cdot (2 - \cos q_x -
\cos q_y)$ with a bandwidth of $B$=1.2~eV and a electron--hole
relative motion continuum, also with a 2d tight binding
dispersion, but a 0.6~eV bandwidth. We have solved numerically Eq.
(\ref{absorption-greens}) with this dispersion law in three cases
of physical interest. In case (1) we consider the situation where
phonon scattering occurs only within the CT-exciton states, i.e.,
CT bound-exciton and its electron--hole continuum, and the
transitions into other exciton bands are not allowed, $\mxd =0$
and $\mcd =0$. In the other cases we assume that phonon scattering
can also occur with other excitonic states\cite{liu93,salamon95}
(LF exciton band) of energy lower than the CT exciton, $\mxd \neq
0$ and $\mcd \neq 0$, and to explore the generality of our
assumptions we consider two slightly different LF exciton
bandwidths in the cases (2a) and (2b).

\begin{figure}
     \begin{center}
        \includegraphics[width=0.85\columnwidth]{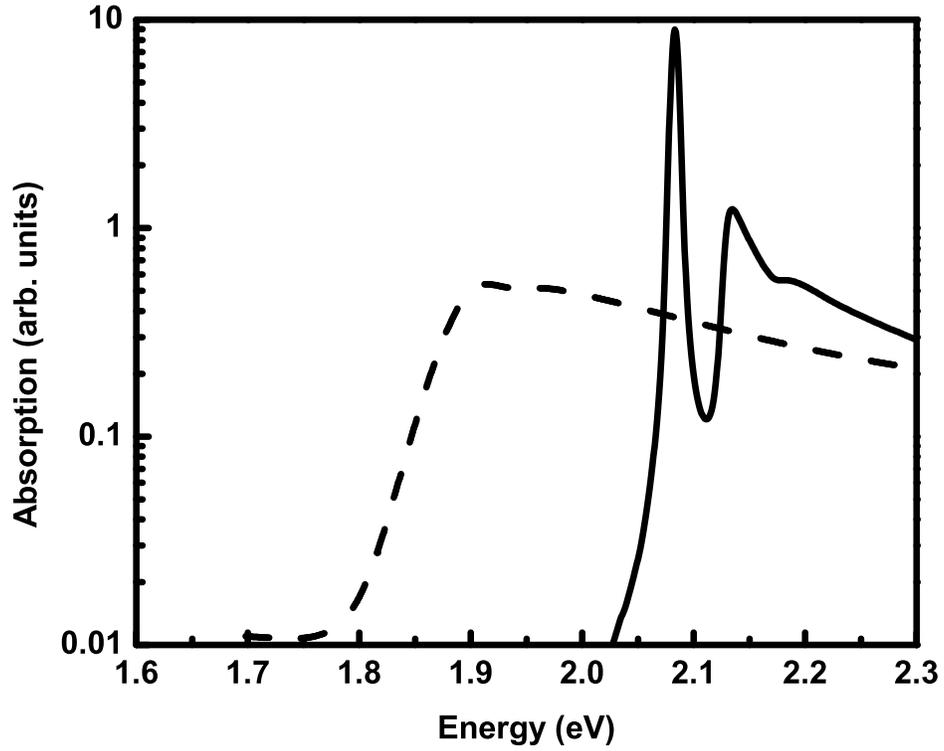}
        \caption{Absorption coefficient for the excitonic contribution
          only on a logarithmic scale, calculated for scattering into
          the exciton and e-h continuum only. Solid line T=15~K, dashed
          line T=300~K. The parameters are the same as for the final
          calculations, except that the coupling constants are twice as
          big and the inhomogenous broadening was only 2 meV.}
         \label{fig:x-without-d}
     \end{center}
\end{figure}

The results in case (1), for the contribution of the bound
CT-exciton ($n=x$ in (\ref{g-resummation})) to the absorption
only, are shown in figure~\ref{fig:x-without-d} and are in obvious
contradiction with experiment. Clearly the coupling to LO--phonons
leads to a broadening of the CT-exciton line. It becomes
asymmetric with a longer tail at the high energy side. On the low
energy side, the spectrum has an Urbach tail, but the width is
extremely small, especially at low temperatures. This is true even
for very high coupling constants. Accounting for the large width
at low temperatures by assuming an homogenous or inhomogenous
broadening would lead to a Lorentzian or Gaussian profile of the
lower energy side of the spectrum.

The shortcoming of our model in case (1) is clear. Since the
Urbach behavior is observed at very low temperature phonon
emission governs the interaction with the lattice as implied by
Eq. (\ref{F1-energy}). Therefore, for broad CT-exciton peak with a
flat exponential tail to occur there must be states available
below its energy and the CT-exciton must be able to couple to
these states by phonons. Only then can LO--phonon scattering be
strong even at zero temperature. Both our own data and the
experiments in Ref.~\onlinecite{choi99} show that such a continuum
exists and starts at about 1.4~eV. It is assigned to Cu d-d
transitions and can couple to the excitons by means of
LO--phonons.

Therefore we expanded the model by including the continuum of LF
excitations. Since little is known about the bandwidth of these
excitations\cite{liu93,salamon95} we have considered two values
for this parameter. In case (2a) we assumed that the LF bandwidth
is the same as that of the CT electron--hole continuum. The CT
exciton binding energy was fitted to be 240~meV, and
$\abs{\coup{}{nm}{}}^2 = 20 (\hbar\omlo)^2$ was used for all the
coupling constants. In addition to the LO--phonon broadening we
included an inhomogeneous broadening $\Gamma_I$=60~meV to account
for sample imperfections. An homogenous broadening also should be
included, however, it must be very small compared to $\Gamma_I$
because otherwise a Lorentzian profile would show up for energies
far below the CT exciton energy in contradiction with experiment.
Therefore, in order to limit the number of parameters we have not
included it in our calculations. Finally, to fit the spectral
contribution of the CT continuum we found that the spectral weight
of the electron--hole continuum needs to be twice that of the
exciton.

\begin{figure}
     \begin{center}
        \includegraphics[width=0.85\columnwidth]{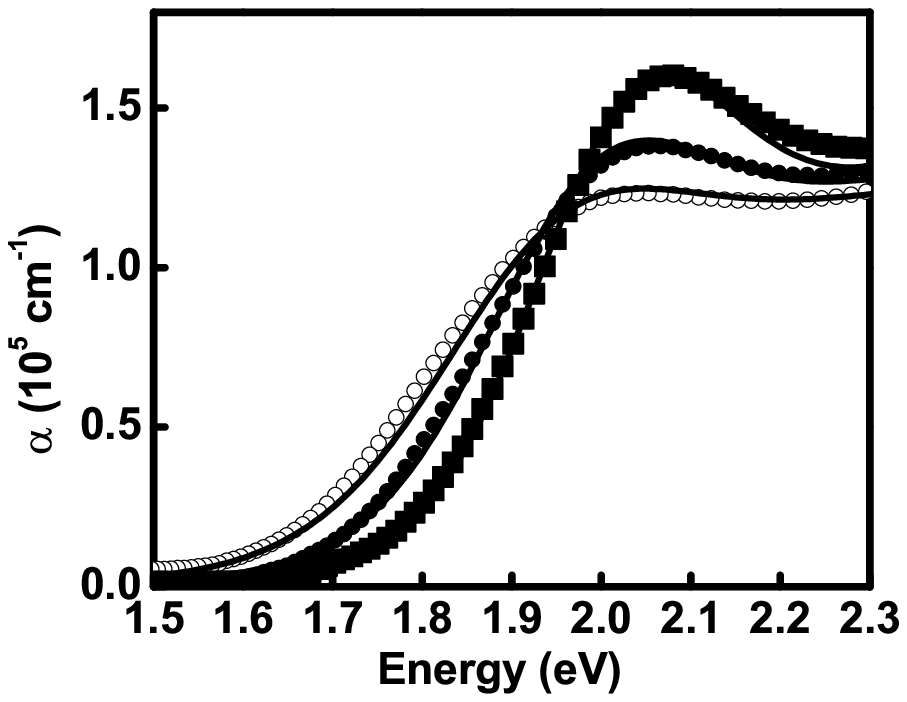}
         \caption{Absorption coefficient data and theory. Only three
         temperatures are shown for clarity. Squares T=15~K, closed
         circles T=250~K, open circles T=350~K, solid lines are the
         theoretical results.}
         \label{theory linear}
     \end{center}
\end{figure}

The solid lines in figures \ref{theory linear} and \ref{theory
log} show the theoretical results in case (2a) both on a linear
and logarithmic scale. They are in excellent agreement with the
experimental data, shown by the dots. The broadening and the shift
of the CT exciton are reproduced correctly and even the decrease
in oscillator strength is well described.

It may be argued that the LF excitations bandwidth should be
smaller than that of the CT electron--hole continuum. In order to
investigate this eventuality, we have considered in case (2b) the
effects of a narrower LF bandwidth, only 1~eV, while maintaining
the flat density of states starting at 1.4~eV in agreement with
the experimental absorption spectrum. In that case we find that we
can also achieve an excellent agreement with experiment, of the
same quality as in case (2a) (not shown), but then we need two
independent coupling constants, one describing the phonon coupling
to the LF excitations and another coupling within the CT manifold.
Interestingly, the numerical integration points out that the
temperature shift of the CT exciton is governed by the coupling to
its own continuum, whereas its broadening is mostly determined by
the coupling to the LF excitations. With these two parameters the
model is also robust against changes of the onset of the LF
excitations.

Although the model formally has a number of fit parameters, many
(phonon energy, CT exciton bandwidths, onset of LF excitations
etc.) are fixed by experiment. Furthermore, we have limited the
number of parameters by limiting the number of phonon coupling
constants to one (case-2a) or two (case-2b), so that in fact only
few parameters are free: the (one or two) phonon coupling
constants, the inhomogenous broadening, the relative spectral
weight of the CT exciton and its continuum, the CT exciton
binding energy and an overall scaling factor.

The results indicate that the coupling constants thus obtained are
indeed large. Due to the different models they cannot be directly
compared to the results of Falck et al. It is straight forward to
check if accounting only for independent one--phonon processes is
sufficient, since one can calculate $\F{2}{}{}{}{}$ with not too
much effort if the $\vq$-dependence of the matrix elements is
neglected. This has not yet been done. Another theory often used
to describe localized electronic excitations coupled to
LO--phonons is Toyozawa's\cite{toyozawa63}. In the strong coupling
regime it predicts a self-trapping of the excitons, i.~e.\  due to
the phonon cloud surrounding the exciton its effective mass
becomes extremely large. Since we have shown on very general
physical grounds that the low temperature Urbach behavior requires
the coupling of the CT-exciton to a continuum of states below the
exciton energy, we expect that a model based on Toyozawa's theory
would give results qualitatively similar to ours.

In conclusion we have presented measurements of the linear
absorption of the material \material\  which exhibits broadening
and shift of the charge transfer gap exciton and, importantly, an
Urbach behavior down to very low temperatures. This behavior
cannot be explained by the current theories. We have developed a
model based on a cumulant expansion that explains well all our
data providing that the charge transfer gap exciton is coupled by
LO--phonons to a continuum of states with lower energy. We
identify this continuum with the Cu d-d continuum which is
revealed in the linear absorption spectra owing to a weak crystal
field symmetry breaking. A consequence of our interpretation is
that the magnetic degrees of freedom do not seem to play a
significant role in the temperature dependence of the charge
transfer absorption edge in this material.

\section{Acknowledgments}
The authors thank W.~Sch\"afer, S.~Louie and A.~Millis for helpful
comments. A.~B.~S. gratefully acknowledges support by the German
National Merit Foundation. The work at Berkeley was supported by
the Director, Office of Science, Office of Basic Energy Science,
Division of Materials Sciences and Office of Science,
U.~S.~Department of Energy under Contract No.~DE-AC03-76SF00098.
The work at Ames Laboratory was supported in part by the DOE
Office of Basic Energy Science under Contract No.~W-7405-Eng-82.


\begin{thebibliography}{10}
\expandafter\ifx\csname bibnamefont\endcsname\relax
  \def\bibnamefont#1{#1}\fi
\expandafter\ifx\csname bibfnamefont\endcsname\relax
  \def\bibfnamefont#1{#1}\fi
\expandafter\ifx\csname url\endcsname\relax
  \def\url#1{\texttt{#1}}\fi
\expandafter\ifx\csname
urlprefix\endcsname\relax\def\urlprefix{URL }\fi
\providecommand{\bibinfo}[2]{#2}

\bibitem[\dagger]{schuh_add}
\bibinfo{note}{Institut f\"ur Angewandte Physik, Universit\"at Karlsruhe, 76128
  Karlsruhe, Germany}.

\bibitem[\ddagger]{dodge_add}
\bibinfo{note}{Present address: Department of Physics, Simon Fraser University,
  Burnaby, BC~V5A~16S, Canada}.

\bibitem{uchida91}
\bibinfo{author}{\bibfnamefont{S.}~\bibnamefont{Uchida}},
  \bibinfo{author}{\bibfnamefont{T.}~\bibnamefont{Ido}},
  \bibinfo{author}{\bibfnamefont{H.}~\bibnamefont{Takagi}},
  \bibinfo{author}{\bibfnamefont{T.}~\bibnamefont{Arima}},
  \bibinfo{author}{\bibfnamefont{Y.}~\bibnamefont{Tokura}}, \bibnamefont{and}
  \bibinfo{author}{\bibfnamefont{S.}~\bibnamefont{Tajima}},
  \bibinfo{journal}{Phys.~Rev.~B} \textbf{\bibinfo{volume}{43}},
  \bibinfo{pages}{7942} (\bibinfo{year}{1991}).

\bibitem{falck92}
\bibinfo{author}{\bibfnamefont{J.~P.} \bibnamefont{Falck}},
  \bibinfo{author}{\bibfnamefont{A.}~\bibnamefont{Levy}},
  \bibinfo{author}{\bibfnamefont{M.~A.} \bibnamefont{Kastner}},
  \bibnamefont{and} \bibinfo{author}{\bibfnamefont{R.~J.}
  \bibnamefont{Birgenau}}, \bibinfo{journal}{Phys.~Rev.~Lett.}
  \textbf{\bibinfo{volume}{69}}(\bibinfo{number}{7}), \bibinfo{pages}{1109}
  (\bibinfo{year}{1992}).

\bibitem{Zhang98}
\bibinfo{author}{\bibfnamefont{F.~C.} \bibnamefont{Zhang}} \bibnamefont{and}
  \bibinfo{author}{\bibfnamefont{K.~K.} \bibnamefont{Ng}},
  \bibinfo{journal}{Phys.~Rev.~B}
  \textbf{\bibinfo{volume}{58}}(\bibinfo{number}{20}), \bibinfo{pages}{13520}
  (\bibinfo{year}{1998}).

\bibitem{tokura90}
\bibinfo{author}{\bibfnamefont{Y.}~\bibnamefont{Tokura}},
  \bibinfo{author}{\bibfnamefont{S.}~\bibnamefont{Koshihara}},
  \bibinfo{author}{\bibfnamefont{T.}~\bibnamefont{Arima}},
  \bibinfo{author}{\bibfnamefont{H.}~\bibnamefont{Takagi}},
  \bibinfo{author}{\bibfnamefont{S.}~\bibnamefont{Ishibashi}},
  \bibinfo{author}{\bibfnamefont{T.}~\bibnamefont{Ido}}, \bibnamefont{and}
  \bibinfo{author}{\bibfnamefont{S.}~\bibnamefont{Uchida}},
  \bibinfo{journal}{Phys.~Rev.~B} \textbf{\bibinfo{volume}{41}},
  \bibinfo{pages}{11657} (\bibinfo{year}{1990}).

\bibitem{zaanen85}
\bibinfo{author}{\bibfnamefont{J.}~\bibnamefont{Zaanen}},
  \bibinfo{author}{\bibfnamefont{G.~A.} \bibnamefont{Sawatzky}},
  \bibnamefont{and} \bibinfo{author}{\bibfnamefont{J.~W.} \bibnamefont{Allen}},
  \bibinfo{journal}{Phys.~Rev.~Lett.} \textbf{\bibinfo{volume}{55}},
  \bibinfo{pages}{418} (\bibinfo{year}{1985}).

\bibitem{choi99}
\bibinfo{author}{\bibfnamefont{H.~S.} \bibnamefont{Choi}},
  \bibinfo{author}{\bibfnamefont{Y.~S.} \bibnamefont{Lee}},
  \bibinfo{author}{\bibfnamefont{T.~W.} \bibnamefont{Noh}},
  \bibinfo{author}{\bibfnamefont{E.~J.} \bibnamefont{Choi}},
  \bibinfo{author}{\bibfnamefont{Y.}~\bibnamefont{Bang}}, \bibnamefont{and}
  \bibinfo{author}{\bibfnamefont{Y.~J.} \bibnamefont{Kim}},
  \bibinfo{journal}{Phys.~Rev.~B}
  \textbf{\bibinfo{volume}{60}}(\bibinfo{number}{7}), \bibinfo{pages}{4646}
  (\bibinfo{year}{1999}).

\bibitem{perkins93}
\bibinfo{author}{\bibfnamefont{J.~D.} \bibnamefont{Perkins}},
  \bibinfo{author}{\bibfnamefont{J.~M.} \bibnamefont{Graybeal}},
  \bibinfo{author}{\bibfnamefont{M.~A.} \bibnamefont{Kastner}},
  \bibinfo{author}{\bibfnamefont{R.~J.} \bibnamefont{Birgeneau}},
  \bibinfo{author}{\bibfnamefont{J.~P.} \bibnamefont{Falck}}, \bibnamefont{and}
  \bibinfo{author}{\bibfnamefont{M.}~\bibnamefont{Greven}},
  \bibinfo{journal}{Phys.~Rev.~Lett.} \textbf{\bibinfo{volume}{71}},
  \bibinfo{pages}{1621} (\bibinfo{year}{1993}).

\bibitem{kuiper98}
\bibinfo{author}{\bibfnamefont{P.}~\bibnamefont{Kuiper}},
  \bibinfo{author}{\bibfnamefont{J.-H.} \bibnamefont{Guo}},
  \bibinfo{author}{\bibfnamefont{C.}~\bibnamefont{Såthe}},
  \bibinfo{author}{\bibfnamefont{L.-C.} \bibnamefont{Duda}},
  \bibinfo{author}{\bibfnamefont{J.}~\bibnamefont{Nordgren}},
  \bibinfo{author}{\bibfnamefont{J.~J.~M.} \bibnamefont{Pothuizen}},
  \bibinfo{author}{\bibfnamefont{F.~M.~F.} \bibnamefont{de~Groot}},
  \bibnamefont{and} \bibinfo{author}{\bibfnamefont{G.~A.}
  \bibnamefont{Sawatzky}}, \bibinfo{journal}{Phys.~Rev.~Lett.}
  \textbf{\bibinfo{volume}{80}}, \bibinfo{pages}{5204} (\bibinfo{year}{1998}).

\bibitem{salamon95}
\bibinfo{author}{\bibfnamefont{D.}~\bibnamefont{Salamon}},
  \bibinfo{author}{\bibfnamefont{R.}~\bibnamefont{Liu}},
  \bibinfo{author}{\bibfnamefont{M.~V. K. M.~A.} \bibnamefont{Karlow}},
  \bibinfo{author}{\bibfnamefont{S.~L.} \bibnamefont{Cooper}},
  \bibinfo{author}{\bibfnamefont{S.}~\bibnamefont{Cheong}},
  \bibinfo{author}{\bibfnamefont{W.~C.} \bibnamefont{Lee}}, \bibnamefont{and}
  \bibinfo{author}{\bibfnamefont{D.~M.} \bibnamefont{Ginsberg}},
  \bibinfo{journal}{Phys.~Rev.~B} \textbf{\bibinfo{volume}{51}},
  \bibinfo{pages}{6617} (\bibinfo{year}{1995}).

\bibitem{wang96}
\bibinfo{author}{\bibfnamefont{Y.~Y.} \bibnamefont{Wang}},
  \bibinfo{author}{\bibfnamefont{F.~C.} \bibnamefont{Zhang}},
  \bibinfo{author}{\bibfnamefont{V.~P.} \bibnamefont{Dravid}},
  \bibinfo{author}{\bibfnamefont{K.~K.} \bibnamefont{Ng}},
  \bibinfo{author}{\bibfnamefont{M.~V.} \bibnamefont{Klein}},
  \bibinfo{author}{\bibfnamefont{S.~E.} \bibnamefont{Schnatterly}},
  \bibnamefont{and} \bibinfo{author}{\bibfnamefont{L.~L.}
  \bibnamefont{Miller}}, \bibinfo{journal}{Phys.~Rev.~Lett.}
  \textbf{\bibinfo{volume}{77}}, \bibinfo{pages}{1809} (\bibinfo{year}{1996}).

\bibitem{Miller90}
\bibinfo{author}{\bibfnamefont{L.~L.} \bibnamefont{Miller}},
  \bibinfo{author}{\bibfnamefont{X.~L.} \bibnamefont{Wang}},
  \bibinfo{author}{\bibfnamefont{S.~X.} \bibnamefont{Wang}},
  \bibinfo{author}{\bibfnamefont{C.}~\bibnamefont{Stassis}},
  \bibinfo{author}{\bibfnamefont{D.~C.} \bibnamefont{Johnston}},
  \bibinfo{author}{\bibfnamefont{J.}~\bibnamefont{{Faber, Jr.}}},
  \bibnamefont{and} \bibinfo{author}{\bibfnamefont{C.}~\bibnamefont{Loong}},
  \bibinfo{journal}{Phys.~Rev.~B} \textbf{\bibinfo{volume}{41}},
  \bibinfo{pages}{1921} (\bibinfo{year}{1990}).

\bibitem{zibold96}
\bibinfo{author}{\bibfnamefont{A.}~\bibnamefont{Zibold}},
  \bibinfo{author}{\bibfnamefont{H.~L.} \bibnamefont{Liu}},
  \bibinfo{author}{\bibfnamefont{S.~W.} \bibnamefont{Moore}},
  \bibinfo{author}{\bibfnamefont{J.~M.} \bibnamefont{Graybeal}},
  \bibnamefont{and} \bibinfo{author}{\bibfnamefont{D.~B.}
  \bibnamefont{Tanner}}, \bibinfo{journal}{Phys. Rev. B}
  \textbf{\bibinfo{volume}{53}}(\bibinfo{number}{17}) (\bibinfo{year}{1996}).

\bibitem{Tajima91}
\bibinfo{author}{\bibfnamefont{S.}~\bibnamefont{Tajima}},
  \bibinfo{author}{\bibfnamefont{T.}~\bibnamefont{Ido}},
  \bibinfo{author}{\bibfnamefont{S.}~\bibnamefont{Ishibashi}},
  \bibinfo{author}{\bibfnamefont{T.}~\bibnamefont{Itoh}},
  \bibinfo{author}{\bibfnamefont{H.}~\bibnamefont{Eisaki}},
  \bibinfo{author}{\bibfnamefont{Y.}~\bibnamefont{Mizuo}},
  \bibinfo{author}{\bibfnamefont{T.}~\bibnamefont{Arima}},
  \bibinfo{author}{\bibfnamefont{H.}~\bibnamefont{Takagi}}, \bibnamefont{and}
  \bibinfo{author}{\bibfnamefont{S.}~\bibnamefont{Uchida}},
  \bibinfo{journal}{Phys.~Rev.~B} \textbf{\bibinfo{volume}{43}},
  \bibinfo{pages}{10496} (\bibinfo{year}{1991}).

\bibitem{kurik71}
\bibinfo{author}{\bibfnamefont{M.~V.} \bibnamefont{Kurik}},
  \bibinfo{journal}{Phys. Stat. Sol. A} \textbf{\bibinfo{volume}{8}},
  \bibinfo{pages}{8} (\bibinfo{year}{1971}).

\bibitem{coupling}
\bibinfo{note}{The definition of the coupling constant in
  Ref.~\onlinecite{falck92} is a factor of 1/2 that of the standard convention
  \cite{mahan}}.

\bibitem{liebler85}
\bibinfo{author}{\bibfnamefont{J.~G.} \bibnamefont{Liebler}},
  \bibinfo{author}{\bibfnamefont{S.}~\bibnamefont{Schmitt-Rink}},
  \bibnamefont{and} \bibinfo{author}{\bibfnamefont{H.}~\bibnamefont{Haug}},
  \bibinfo{journal}{J.~Lum.} \textbf{\bibinfo{volume}{34}}, \bibinfo{pages}{1}
  (\bibinfo{year}{1985}).

\bibitem{liu93}
\bibinfo{author}{\bibfnamefont{R.}~\bibnamefont{Liu}},
  \bibinfo{author}{\bibfnamefont{D.}~\bibnamefont{Salamon}},
  \bibinfo{author}{\bibfnamefont{M.~V.} \bibnamefont{Klein}},
  \bibinfo{author}{\bibfnamefont{S.~L.} \bibnamefont{Cooper}},
  \bibinfo{author}{\bibfnamefont{W.~C.} \bibnamefont{Lee}},
  \bibinfo{author}{\bibfnamefont{S.}~\bibnamefont{Cheong}}, \bibnamefont{and}
  \bibinfo{author}{\bibfnamefont{D.~M.} \bibnamefont{Ginsberg}},
  \bibinfo{journal}{Phys.~Rev.~Lett.} \textbf{\bibinfo{volume}{71}},
  \bibinfo{pages}{3709} (\bibinfo{year}{1993}).

\bibitem{toyozawa63}
\bibinfo{author}{\bibfnamefont{Y.}~\bibnamefont{Toyozawa}},
  \emph{\bibinfo{title}{Polarons and Excitons}} (\bibinfo{publisher}{Plenum},
  \bibinfo{address}{New York}, \bibinfo{year}{1963}).

\bibitem{mahan}
\bibinfo{author}{\bibfnamefont{G.~D.} \bibnamefont{Mahan}},
  \emph{\bibinfo{title}{Many--particle physics}} (\bibinfo{publisher}{Plenum
  Pub Corp.}, \bibinfo{year}{1990}), \bibinfo{edition}{2nd} ed.

\end{thebibliography}

\begin{figure}
     \begin{center}
         \includegraphics[width=0.85\columnwidth]{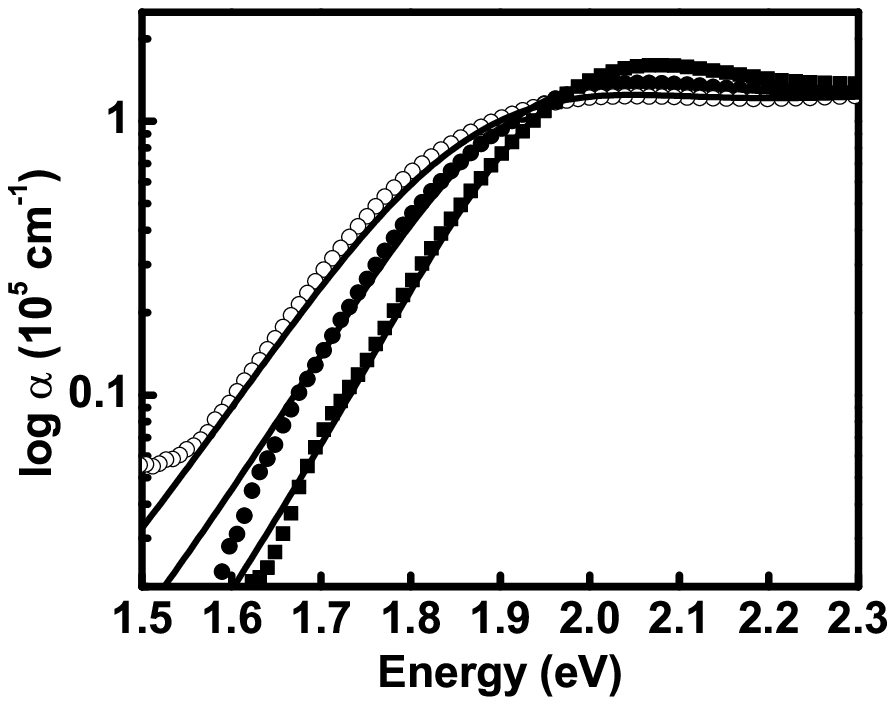}
         \caption{Absorption coefficient data and theory, displayed
         on a semi--logarithmic scale. Only three temperatures are
         shown for clarity. Squares T=15~K, closed circles T=250~K,
         open circles T=350~K, solid lines are the theoretical
         results.}
         \label{theory log}
     \end{center}
\end{figure}

\end{document}